\documentclass[aps,prl,twocolumn,superscriptaddress]{revtex4-1}
\usepackage{amsmath, amsthm, amssymb,latexsym,mathrsfs,dsfont}    
\usepackage{graphicx}
\usepackage{xcolor}
\usepackage{color}
\usepackage[caption=false]{subfig} 
\usepackage{braket}
\usepackage{cancel}
\usepackage{array}
\usepackage{slashed}
\usepackage{bookmark,bbm}
\newcommand{\vect}[1]{\boldsymbol{#1}} 
\usepackage{cancel}
\graphicspath{{./Figs/}}
\newcommand{\dd}{\mathrm{d}}

\newcommand{\lp}{{^{+}}}

\newcolumntype{P}[1]{>{\centering\arraybackslash}p{#1}}

\hypersetup{pdftitle={main.pdf},hidelinks
}
\begin{document}
	
	
	\title{Towards solving the proton spin puzzle}
	
	\author{Andreas Ekstedt}
	\email{andreas.ekstedt@physics.uu.se}
	\affiliation{Department of Physics and Astronomy, Uppsala University, Box 516, SE-751 20 Uppsala, Sweden}
	
	\author{Hazhar Ghaderi}
	\email{hazhar.ghaderi@physics.uu.se}
	\affiliation{Department of Physics and Astronomy, Uppsala University, Box 516, SE-751 20 Uppsala, Sweden}
	
	\author{Gunnar Ingelman}
	
	\email{gunnar.ingelman@physics.uu.se}
	
	\affiliation{Department of Physics and Astronomy, Uppsala University, Box 516, SE-751 20 Uppsala, Sweden}
	
	\affiliation{Swedish Collegium for Advanced Study, Thunbergsv\"agen 2, SE-752 38 Uppsala, Sweden}
	
	\author{Stefan Leupold}
	\email{stefan.leupold@physics.uu.se}
	\affiliation{Department of Physics and Astronomy, Uppsala University, Box 516, SE-751 20 Uppsala, Sweden}
	
	\date{June 19, 2019 }
	
	\begin{abstract} 
		The fact that the spins of the quarks in the proton, as measured in deep inelastic lepton-proton scattering, only add up to about 30\% of the spin of the proton is still not understood after 30 years. 
		We show that our newly developed model for the quark and gluon momentum distributions in the proton, based on quantum fluctuations of the proton into baryon-meson pairs convoluted with Gaussian momentum distributions of partons in hadrons, can essentially reproduce the data on the proton spin structure function $g_1^P(x)$ and the associated spin asymmetry. A further improved description of the data is achieved by also including the relativistic correction of the Melosh transformation to the light-front formalism used in deep inelastic scattering. However, this does not fully resolve the spin puzzle, including also the neutron spin structure and the spin sum rules. These aspects can also be accounted for by our few-parameter model if the conventional SU(6) flavor-spin symmetry is broken, giving new information on the non-perturbative bound-state nucleon. 
	\end{abstract}
	
	\maketitle

	\section{Introduction}\label{sec:Introduction}
	The EMC experiment \cite{Ashman:1987hv} at CERN initiated the proton spin puzzle through a measurement showing that the spins of the quarks in a proton only add up to a fraction of the spin of the proton. This triggered large efforts on both the experimental and theoretical side to investigate the problem and interpret the data, see the reviews \cite{JAFFE1990509, Filippone:2001ux, Myhrer:2009uq, Aidala:2012mv,Bass:2009dr}. In spite of substantial advances giving much more data and more elaborate theoretical calculations the basic problem remains after 30 years. The spin 1/2 of the proton (along a quantization axis) is given by the sum of the angular momenta of its constituents 
	\begin{equation}\label{E: spin-decomposition}
	{1}/{2}= \Delta\Sigma/2 + \Delta g + L_q + L_g ,
	\end{equation}
	where $ \Delta\Sigma=\sum_{q=u,d,s...} \Delta q$ sums the difference $\Delta q = q^\uparrow - q^\downarrow$ of spin along and opposite the proton spin carried by all quarks and antiquarks in the proton. Similarly $\Delta g = g^\uparrow - g^\downarrow$ for the spin of gluons, whereas $L_q$ and  $L_g$ denote the possible orbital angular momenta carried by quarks or gluons. The measurements \cite{Adolph:2015saz,Airapetian:2006vy} give $\Delta\Sigma\sim 0.3$ and $\Delta g$ so small that a significant part of the proton spin remains unexplained. 
	
	Deep inelastic scattering (DIS) of electrons or muons on protons probes the proton  structure by the exchange of a virtual photon (four-momentum $q$, momentum transfer $Q^2=-q^2$). For unpolarized protons the inclusive cross-section is $\dd\sigma/(\dd x\,\dd Q^2) \sim F_2(x,Q^2)$, where the proton structure function
	\begin{equation}\label{E: F2}
	F_2(x,Q^2)=\sum\nolimits_q e_q^2 x\left( q(x,Q^2) + \bar{q}(x,Q^2) \right),
	\end{equation}
	is interpreted in terms of parton distribution functions (PDFs) $q(x,Q^2)$ for the probability to find a quark, of flavor $q$ and charge $e_q$, carrying energy-momentum fraction $x$ of the proton when probed at the scale $Q^2$, and analogously for antiquarks $\bar{q}$. For polarized protons, the polarized structure function is defined as
	\begin{equation}\label{E: g1}
	g_1(x,Q^2)=\frac{1}{2}\sum\nolimits_q e_q^2\Delta q(x,Q^2),
	\end{equation}
	in terms of polarized PDFs $\Delta q(x,Q^2)=q^{\uparrow}(x,Q^2)-q^{\downarrow}(x,Q^2)$ such that the above $\Delta q = \int_0^1 \dd x \Delta q(x,Q^2) $.
	
	The unpolarized quark and gluon PDFs are well constrained from structure function measurements, with $Q^2$-dependence well understood in terms of the DGLAP equations \cite{Gribov:1972ri,Dokshitzer:1977sg,Altarelli:1977zs} from perturbative QCD (pQCD). However, the basic $x$-dependence at the starting scale $Q_0^2\sim 1~\text{GeV}^2$ of pQCD evolution originates from the bound-state proton where no proper theoretical description is available for its soft non-perturbative QCD dynamics. PDFs are, therefore, conventionally described by many-parameter fits without physics insights. We have recently presented \cite{Ekstedt:2018onk} a physically motivated, few-parameter, model for the PDFs at $Q_0^2$, which is here used to study the spin degree of freedom and thereby the proton spin puzzle.

	\section{Spin-dependent parton distributions}\label{sec:SPDF}
	
	Our model for the PDFs in the proton is a convolution of quantum fluctuations at the hadron level, and at the parton level. The essential parts are reviewed here, referring to \cite{Ekstedt:2018onk} and references therein for details. 
	At the hadron level, the proton can fluctuate into baryon-meson ($BM$) states, giving the proton quantum state  
	\begin{eqnarray}
	\ket{P} = \alpha_{\rm bare} \ket{P}_{\rm bare} + \sum\nolimits_{BM} \alpha_{BM} \ket{BM}
	\label{E: FockExpansion}
	\end{eqnarray}
	with $BM=N\pi,\Delta\pi,\Lambda K^+,\cdots$.
	Using the well established leading-order Lagrangian of three-flavor chiral perturbation theory \cite{Jenkins:1991es, Pascalutsa:1999zz, Pascalutsa:2006up, Ledwig:2014rfa}, our model gives the probabilities $|\alpha_{BM}|^2$ for the different fluctuations, the distribution of their internal momenta and includes the spin degree of freedom. For a detailed account of the chiral symmetry basis for the baryon-meson Fock components we refer to \cite{Arndt:2001ye,Chen:2001eg,Chen:2001pva,Thomas:2000ny,Ji:2013bca,Wang:2016ndh}. The different terms are theoretically well defined and related to each other with only three coupling constants known from hadronic processes and weak decays of baryons \cite{Ekstedt:2018onk}.
	
\begin{figure}[!t]
		\centering
		\subfloat[]{\label{Fig: dis_bare}\includegraphics[width=0.33\columnwidth]{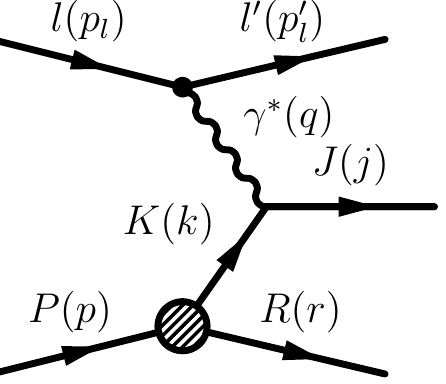}}
			\subfloat[]{\label{Fig: dis_meson}\includegraphics[width=0.33\columnwidth]{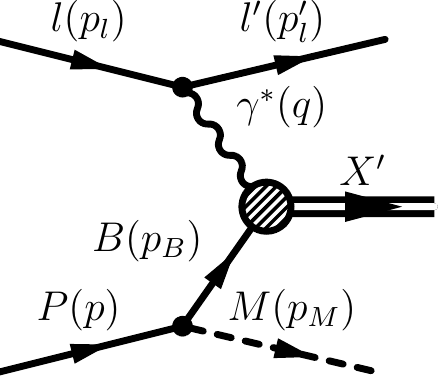}}
		\subfloat[]{\label{Fig: dis_baryon}\includegraphics[width=0.33\columnwidth]{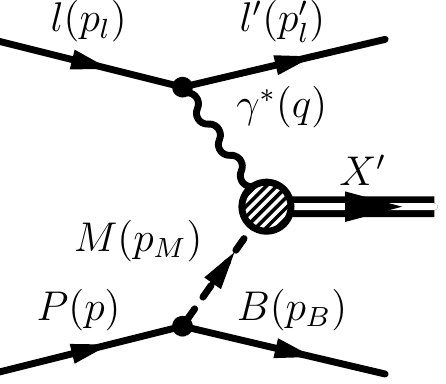}}
		\caption{Lepton scattering, via virtual photon exchange, on a quark in (a) the bare proton, (b) the baryon or (c) the meson in a baryon-meson quantum fluctuation of the proton. }
		\label{Fig: DISfluc}
	\end{figure}

	The importance of these hadronic fluctuations for the spin measurement is immediately clear from the fact that the nucleon-pion is in a p-wave due to the negative parity of the pion. This orbital angular momentum of the whole hadronic system is not observable by a point-like large-$Q^2$ photon that couples to a single quark. Moreover, the nucleon in the fluctuation can have its spin flipped compared to the original proton spin and thereby a probed quark in the nucleon can give a contribution of opposite polarization. A probed quark in the spin-zero pion will give zero contribution to the measured polarization. Thus, the hadronic fluctuations reduce the measured polarization at the quark level. To account for this in DIS we allow for the possibilities that the virtual photon probes a parton either in the bare proton, or in the meson ($M$) or the baryon ($B$) of a fluctuation, see Fig.\ \ref{Fig: DISfluc}.
	
	Using light-cone time-ordered perturbation theory, we obtain the probability distribution as a function of the light-cone fraction $y=p_B^+/p_P^+$ for a probed baryon with helicity $\lambda$ in a baryon-meson fluctuation (for the proton polarized along the $+\hat{z}$ axis)
	\begin{equation}
	\label{E: HadDist}
	f^{\lambda}_{BM}(y) \! = \! \frac{|g_{BM}|^2}	
	{ 2y(1-y)}\int\! \frac{\dd^2\vect{k}_\perp}{(2\pi)^3} \left|	
	\frac{G(y,k_\perp^2,\Lambda_H^2)
		S^{\lambda}\left(y,\vect{k}_\perp\right)}{m_P^2-m^2(y,k_\perp^2)}\right|^2\!. 
	\end{equation}  
	The corresponding function for probing the meson is
	\begin{equation}\label{E: MesonDist} 
	f_{MB}(y) = f_{BM}(1-y) \equiv \sum\nolimits_\lambda f_{BM}^\lambda(1-y).
	\end{equation}
	The propagator encodes the suppression due to the virtuality of the fluctuation given by the difference of the squared masses of the proton and the baryon-meson system $m^2(y,k_\perp^2)\equiv (m_B^2+k_\perp^2)/y + (m_M^2+k_\perp^2)/(1-y)$.
	
	Ref.\ \cite{Ekstedt:2018onk} specifies the details of the hadronic couplings $g_{BM}$, the vertex functions $S^{\lambda}\left(y,\vect{k}_\perp\right)$ and the cut-off form factor $G\left(y,k_\perp^2,\Lambda_H^2\right) = \exp\left[-\mathcal{A}^2/( 2\Lambda_H^2)\right] $. 
	The form factor accounts for the fact that the description in terms of hadronic degrees of freedom is only valid at hadronic scales and therefore exponentially suppressed for hadronic momentum transfers above the scale parameter $\Lambda_H$. Because $\Lambda_H$ is related to the switch to partonic degrees of freedom, one expects $\Lambda_H$ to be of the same order as the starting scale $Q_0$ of the pQCD formalism. The fluctuation probabilities in (\ref{E: FockExpansion}) are
	$\lvert \alpha_{BM}(\Lambda_H)\rvert^2  = \int_0^1\! \dd y\,  f_{BM}  (y)$ giving an overall fluctuation probability of several tens of percent \cite{Ekstedt:2018onk}.
	
	Due to the hadronic fluctuations, PDFs for the unpolarized proton are given by a bare part and a convolution part 
	\begin{align}\label{E: model-PDF}
	&~f_{i/P}(x) =  f_{i/P}^{\text{bare}}(x) \nonumber
	\\ 
	&
	+ \sum\nolimits_{\lambda, H\in\{B,M\}}
	\int\!  \dd y\, \dd z \,\delta(x-y z)f^{\text{bare}}_{i/H}(z) f_{H/P}^{\lambda} (y), 
	\end{align}
	with the hadronic distributions $f_{H/P}(y)$ given by (\ref{E: HadDist},\ref{E: MesonDist}). 
	
	Our model \cite{Ekstedt:2018onk} for the bare PDFs at the starting scale $Q_0^2$ includes valence quarks and gluons, whereas sea quarks are generated by our hadronic fluctuations. 
	
	In the hadron rest frame, with no preferred direction, it is natural to assume a spherically symmetric momentum distribution. In the absence of a proper description derived from QCD we assume a Gaussian shape, that suppresses large momentum fluctuations, and may represent the added effect of many soft momentum exchanges within the bound state hadron.
	The four-momentum distribution for a parton of type $i=q, \bar{q}, g$ of mass $m_i$ in hadron $H$ is therefore given by
	\begin{equation}\label{E: FH}
	F_{i/H}(k) 
	=
	N_{i/H} 
	\exp{\left[-\frac{(k_0-m_i)^2+k_x^2+k_y^2+k_z^2}{2\sigma_i^2}\right]},
	\end{equation}
	where $N$ is a normalization factor. The width $\sigma$ is expected to be physically given by the uncertainty relation $\Delta x\Delta p \sim \hbar/2$ that enforces increasing momentum fluctuations for a particle confined in a smaller spatial range. With the hadron size as $\Delta x$ one expects $\sigma \sim 0.1\,$GeV. 
	
	Using light-cone momenta, $x={k^+}/{p_H^+}$ is the proper energy-momentum fraction (independent of longitudinal boosts) carried by a parton in a hadron, giving \cite{Ekstedt:2018onk}
	\begin{align}\label{E: PDF-unpol}
	f^{\text{bare}}_{i/H}(x) = \int' \!\frac{\dd^4k}{(2\pi)^4}\, 
	\delta\left({k^+}/{p_H^+}-x\right)F_{i/H}(k). 
	\end{align}
	The prime on the integral sign indicates the kinematical constraints on the $k$-integral: the scattered parton four-vector $j$ must be on-shell or have a time-like virtuality (causing final-state QCD radiation) limited by the mass of the hadronic system, i.e.\ $m_i^2 < j^2 = (k+q)^2 < (p_H+q)^2$, and the hadron remnant must have a four-vector $r^2=(p_H-k)^2>0$, cf.\ Fig.\ \ref{Fig: dis_bare}. 
		This introduces the four-momentum $q$ of the virtual photon which reshapes the originally spherically symmetric function (\ref{E: FH}) into a distribution that contains the directional information from the  probe of the DIS measurement process. The constraints also imply that $x$ is restricted to its physical range $0<x<1$, and that the PDFs vanish smoothly as $x\to 1$ \cite{Ekstedt:2018onk}.
	
	The normalization factors $N_{q/H}(\sigma_q,m_q)$ are fixed by the flavor sum rules, that is\ $N_i=\int_0^1\! \dd x\, f^{\text{bare}}_{i/H}(x)$\textemdash giving the correct number ($N_i$) of valence quarks of flavor $i$ in a given hadron $H$. The gluon normalization $N_{g/H}(\sigma_g,0)$ is fixed by momentum conservation, that is, by the requirement \ $\sum_{i} \int_{0}^{1}\! \dd x\; xf^{\text{bare}}_{i/H}(x)=1$.
	
	The Gaussian widths are taken as $\sigma_g$, $\sigma_1$, and $\sigma_2$ for the distributions of the gluon and for a quark flavor occurring once ($\int_0^1\! \dd x\, f^{\text{bare}}_{q/H}(x) =1$) or twice ($\int_0^1\! \dd x \, f^{\text{bare}}_{q/H}(x)=2$) in a hadron. For instance $\sigma_2$ ($\sigma_1$) is the Gaussian width for the $u$ ($d$) distribution in the proton. The distributions containing three quarks of the same flavor can be obtained using isospin relations (throughout this work we ignore isospin breaking effects, which are generically on the order of 1\%). 
	
	With this $f^{\text{bare}}_{i/H}(x)$ of a hadron inserted in (\ref{E: model-PDF}) the PDFs of the proton quantum state are obtained at $Q_0^2$ and then evolved to larger $Q^2$ with conventional DGLAP equations \cite{Botje:2010ay}. The values of the five free parameters in the model are obtained by fitting proton structure function data from inclusive DIS on unpolarized protons \cite{Ekstedt:2018onk}:
	\begin{align}\label{E: parameter-values}
	&\sigma_1=0.11~\text{GeV},~\sigma_2=0.22~\text{GeV},~\sigma_g=0.028~\text{GeV},  
	\nonumber
	\\
	&Q_0=0.88~\text{GeV},~\Lambda_H=0.87~\text{GeV}. 
	\end{align}
	It is noteworthy that the Gaussian widths have reasonable magnitudes \cite{Ekstedt:2018onk} and that $\Lambda_H$ and $Q_0$ have essentially the same value, as expected because both relate to the transition from hadron to parton degrees of freedom.
	
	As basis for the spin dependent PDFs,  $\Delta f_i(x)=f_i^{\uparrow}(x)-f_i^{\downarrow}(x)$, at $Q^2_0$ it is suggestive to use as a first approximation the non-relativistic SU(6) quark model, combining the global $\mathrm{SU}(3)$ flavor symmetry $(u,d,s)$ with the $\mathrm{SU}(2)$ spin group. This results in well defined spin and flavor decomposition of all used baryons and mesons \cite{mosel1999fields}, e.g.\ for the proton giving $\Delta f_u^P = {4}/{3}$ and $\Delta f_d^P=- {1}/{3}$. We use such $\Delta f$ from $\mathrm{SU}(6)$ to rescale the unpolarized $f^{\text{bare}}_{i/H}(x)$ of (\ref{E: PDF-unpol}) by the factor $\Delta f_{i/H}^\text{SU(6)}=\Delta f_{i/H} / \int_0^1\! \dd z\, f^{\text{bare}}_{i/H}(z)$ to obtain its spin-dependent version with correct normalization. 
	
	Concerning the spins of the partons in a hadron, the discussion has so far been limited to a non-relativistic treatment where spin is treated independently from motion. However, DIS probes the proton on the light-cone, a highly relativistic process. The partons move relative to the hadron, so their helictites do not necessarily agree with their spin component along the spin quantization axis of the proton (which is conveniently chosen along the collision axis of the proton and the virtual photon). The effect of the probed quark's transverse motion on the spin measurement can be accounted for by the Melosh transformation, leading to the suppression factor \cite{Melosh:1974cu, Ma:1991xq, Beyer:1998xy, Sun:2001ir, SCHMIDT1997331}
	\begin{equation}\label{E: Melosh}
	M(k) = \frac{(k^+ +m_q)^2-k_\perp^2}{(k^{+}
		+m_q)^2+k_\perp^2}.
	\end{equation}
	Including this suppression factor when integrating over the momentum $k$ in Eq.\ (\ref{E: PDF-unpol}), we obtain the expression for the polarized PDFs of a bare hadron 
	\begin{align}\label{E: Melosh-PDF}
	\Delta f^{\text{bare}}_{i/H}(x) =& \Delta f_{i/H}^\text{SU(6)} \int' \!\frac{\dd^4k}{(2\pi)^4}\, 
	\delta\left(\frac{k\lp}{p_H^+}-x\right) F_{i/H}(k) M(k)
	\end{align}
	where the integral is subject to the same energy-momentum constraints as for the unpolarized case in Eq.\ (\ref{E: PDF-unpol}).
	To include also the hadronic fluctuations, the full $\Delta f_i$ is obtained from (\ref{E: model-PDF}) by replacing $f^\text{bare}_{i/H}(z)$ therein by $\Delta f^{\text{bare}}_{i/H}(z)$ from (\ref{E: Melosh-PDF}) and properly accounting for the spin dependence of the hadronic vertex functions $S^{\lambda}(y,\vect{k}_\perp)$.
	
		The Melosh factor (\ref{E: Melosh}) is based on the relativistic formalism with on-shell four-momenta; off-shell partons do not have a well-defined Melosh transformation although one may argue that time-like momenta should behave in a similar manner. 
		In our confinement-based model, Eq.\ (\ref{E: FH}), with Gaussian fluctuations of order 0.1 GeV of all components of a parton's four-vector $k$, the probed quark will typically be somewhat off-shell. The effective quark mass can therefore be of order 0.1 GeV, even for bare quark masses of a few MeV, as used in (\ref{E: FH}). It is then natural to replace $m_q$ in the Melosh factor (\ref{E: Melosh}) by $\sqrt{|k^2|}$, where the modulus is to also include fluctuations with a slightly space-like $k$-vector. The Melosh transformation then results in an overall relativistic spin suppression factor of 0.52 for the integrated spin PDFs of quarks. This factor becomes 0.7 when space-like $k$-vectors of the probed quark are rejected.

\begin{figure}[h]
		\includegraphics[width=0.5\textwidth]{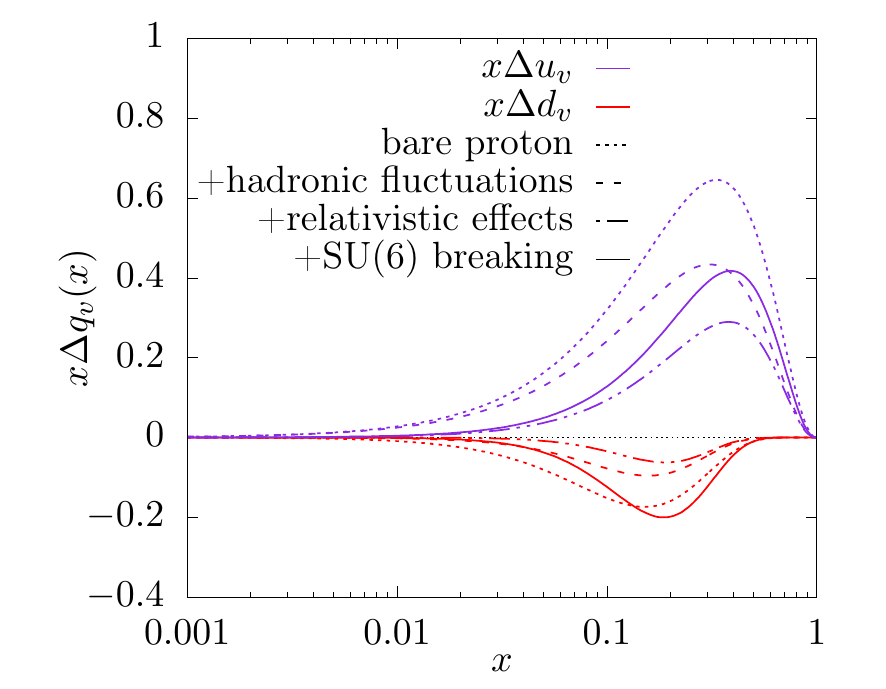}
		\caption{Spin dependent proton PDFs for $x\Delta u_v=x\Delta u-x\Delta \bar{u}$ and $x\Delta d_v=x\Delta d-x\Delta \bar{d}$.
			Model curves for bare proton (dotted), and successively adding the effects of hadronic fluctuations (dashed), relativistic Melosh transformation (dash-dotted) and $\mathrm{SU}(6)$ breaking (full).}
		\label{figure:PDFS}
	\end{figure}

		As seen from Eq.\ (\ref{E: Melosh}) the main effect of the relativistic Melosh correction arises from the transverse momentum $k_{\perp}^2$ of the probed quark. The net effect therefore is not only an overall suppression, but also a stronger suppression at lower $x$ and therefore a somewhat modified $x$-shape of the spin PDFs relative to the unpolarized PDFs. Comparing in Fig.\ \ref{figure:PDFS} the dashed to the dash-dotted curves illustrates this suppression and the shift of the respective peak to larger $x$ values.
	
	\section{Comparison to spin structure data}\label{Sec: Data-comparison}
	
\begin{figure}
		\centering
		\includegraphics[width=1\columnwidth]{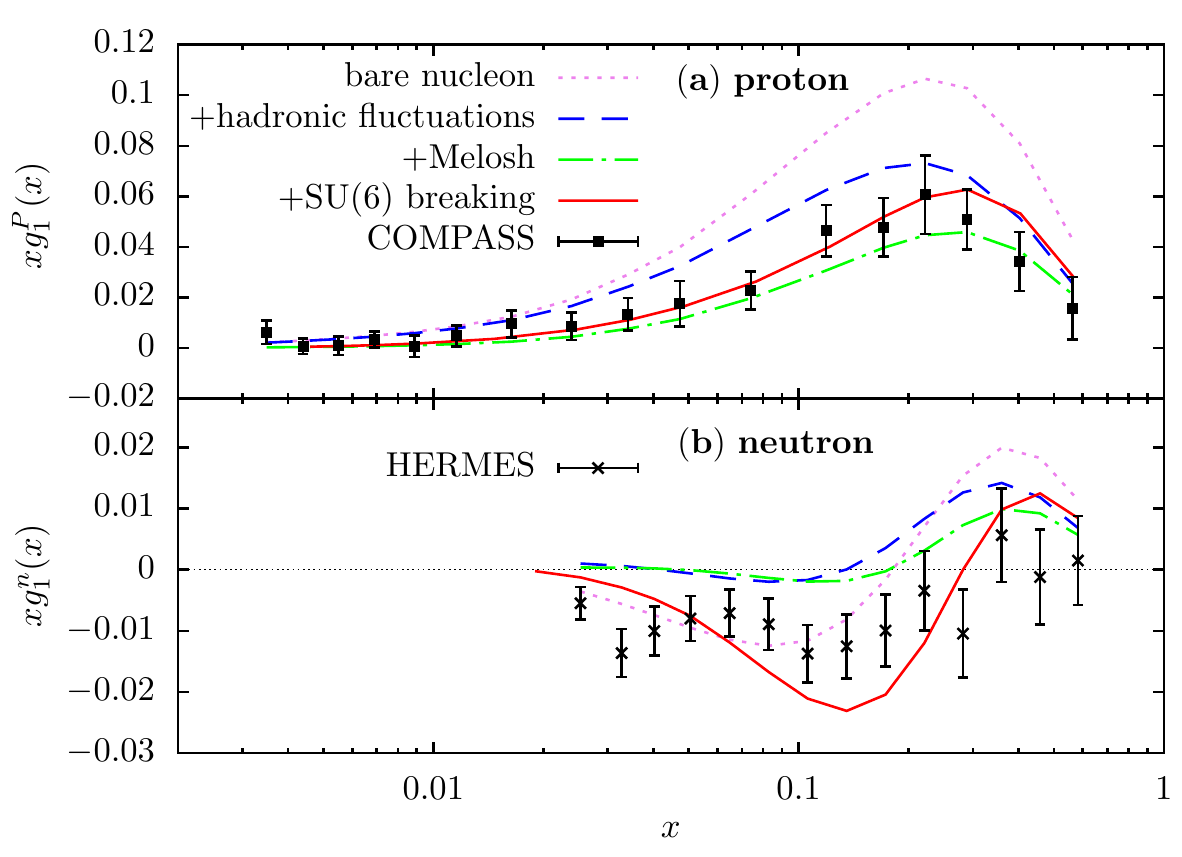}
		\caption{The spin structure function $xg_1(x)$ of (\ref{E: g1}) for quarks (a) in the proton with data from COMPASS \cite{Adolph:2015saz} and (b) in the neutron with data from HERMES \cite{Airapetian:2006vy}.  Model curves for bare nucleon (dotted), and successively adding the effects of hadronic fluctuations (dashed), relativistic Melosh transformation (dash-dotted) and $\mathrm{SU}(6)$ breaking (full).} 
		\label{Fig: g1x}
	\end{figure}

\begin{figure}
		\centering
		\includegraphics[width=1\columnwidth]{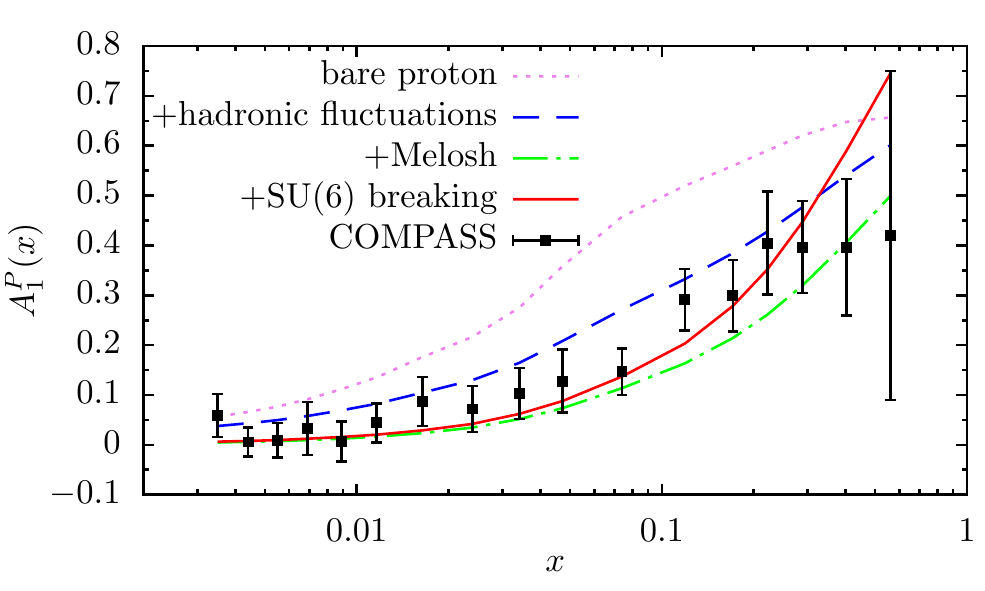}
		\caption{The proton spin asymmetry $A_1^P(x)$. Data from COMPASS \cite{Adolph:2015saz} and model curves as in Fig.\ \ref{Fig: g1x}.} 
		\label{Fig: A1}
	\end{figure}

\begin{figure}
		\centering
		\includegraphics[width=1\columnwidth]{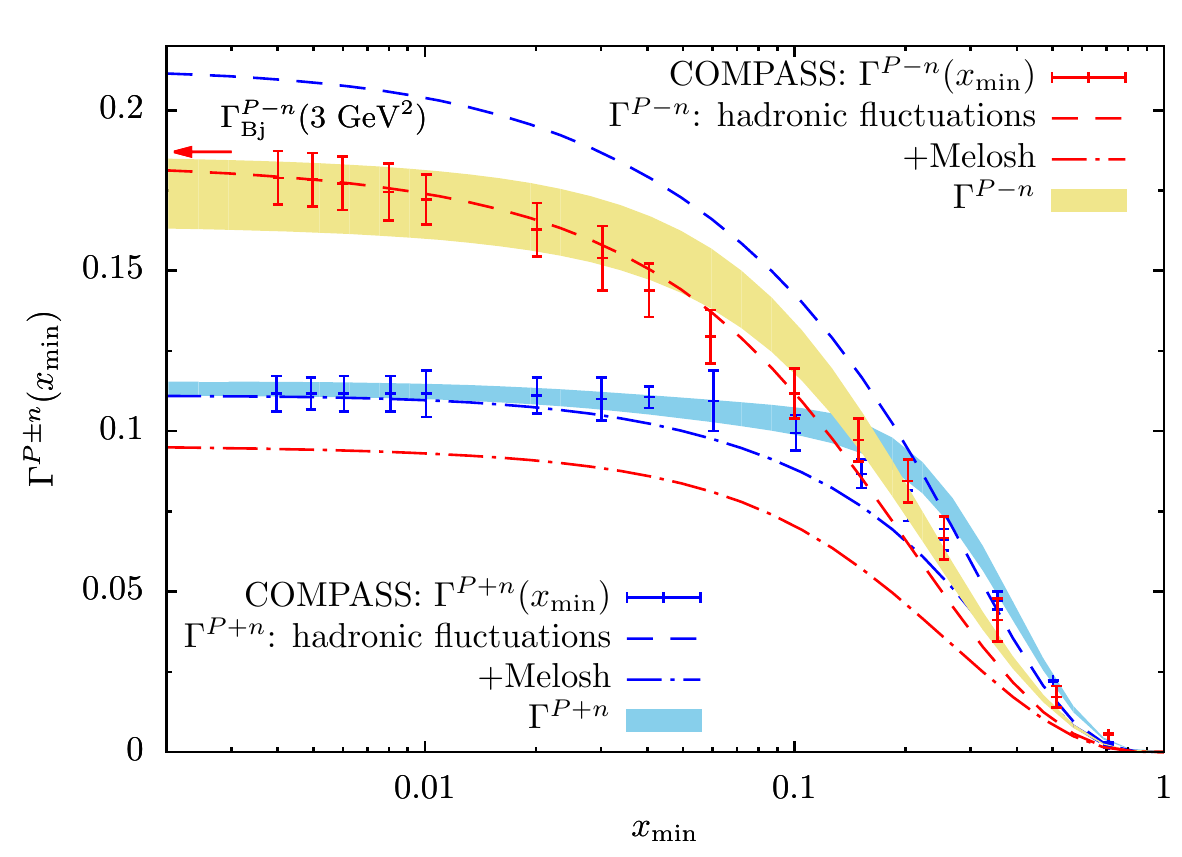}
		\caption{The integrals $\Gamma^{P\pm n}(x_\text{min}) = \int_{x_\text{min}}^{1}\!\dd x\, \left(g_1^P\pm g_1^n \right)$ as a function of $x_\text{min}$.  
			Data from COMPASS \cite{Adolph:2015saz,Alexakhin:2006oza,Aidala:2012mv} and model curves for nucleon with hadronic fluctuations (dashed), including relativistic Melosh transformation (dash-dotted) and also including $\mathrm{SU}(6)$ breaking (bands).} 
		\label{Fig: sumrules}
	\end{figure}

	Since the five parameters (\ref{E: parameter-values}) in our model are determined from unpolarized structure function data, there is no remaining free parameter to fit to measured spin structure functions. When comparing to data for polarized protons and neutrons in Figs.\ \ref{Fig: g1x} to \ref{Fig: sumrules} and in Table \ref{Table: DeltaSigma}, we show the effect of including the different ingredients in the model, first only the bare nucleon, then adding hadronic fluctuations and then also the relativistic correction via the Melosh transformation. 
	
	Considering first the proton spin structure function $xg_1(x)$ in Fig.\ \ref{Fig: g1x}a it is clear that having only the bare proton gives a too large fraction of the proton spin carried by the quarks, but including the hadronic fluctuations lowers the result substantially to be systematically just slightly above data. This shows that a large part of the missing spin is related to the effects discussed above in terms of orbital angular momentum of the baryon-meson system and contributions of spin-flipped baryon and zero-spin pion. Adding also the Melosh transformation the resulting $xg_1(x)$ is further reduced to be slightly on the low side of the data. From Fig.\ \ref{Fig: A1} similar conclusions are obtained by the comparison of our model to the measured proton spin asymmetry which is essentially the ratio of the polarized and the unpolarized structure function, $A_1(x)=\left[g_1(x)-\frac{Q^2}{\nu^2}g_2(x)\right]/F_1(x)\approx 2xg_1(x)/F_2(x)$ \cite{Aidala:2012mv}.
	
	Including these spin structure function data in a fine-tuning of the model parameters one may well get an even better description of these data. Considering only the proton spin structure one could then be satisfied and consider the proton spin puzzle solved. However, there are more pieces in this puzzle, namely the neutron spin and the spin sum rules.

	The spin structure function of the neutron,  $xg^{n}_1(x)$ in Fig.\ \ref{Fig: g1x}b, has much smaller magnitude than that of the proton. This is expected because with the bare neutron state given by an isospin flip, meaning $\mathrm{SU}(6)$ spin factors $\Delta f_d^n=\Delta f_u^P=4/3$, $\Delta f_u^n=\Delta f_d^P=-1/3$ 
	and PDFs $d_n(x)=u_P(x)$, $u_n(x)=d_P(x)$ and assuming the same $x$-shape of $u(x)$ and $d(x)$, then the balance of squared quark charges and $\Delta q$'s gives  $xg^{n}_1(x)=0$. This is close to being the case, but not quite. In our model the $x$-shapes are slightly different due to $\sigma_2>\sigma_1$, meaning that the neutron's $d$-distribution is harder, but the small $d$-charge makes $xg_1^n$ only slightly positive at large $x$, whereas the four times larger charge-squared of the $u$-quark with its softer $x$-shape gives negative $xg_1^n$ at smaller $x$. The data on $xg^{n}_1(x)$ show this trend qualitatively. 
	
	More information is available via the integrals \cite{Ellis:1973kp,Bjorken:1966jh,PhysRevD.1.1376}
	\begin{equation}\label{E: Sumrule-integrals}
	\Gamma^{P\pm n}(x_\text{min}) = \int_{x_\text{min}}^{1} \! \dd x \, 
	\left(g_1^P(x)\pm g_1^n(x) \right)
	\end{equation}
	shown in Fig.\ \ref{Fig: sumrules} for data and model. The naive model with only the bare nucleons gives too high values which are not shown. The data on $\Gamma^{P+n}$ can be qualitatively described when including both hadronic fluctuations and the Melosh transformation, although being below data in the valence quark $x$-region. The data on $\Gamma^{P-n}$ are rather well described by the model including hadronic fluctuations, but the model falls far below the data when including the Melosh transformation. 

	The integral $\Gamma^{P-n}(0)$ constitutes the Bjorken sum rule \cite{Bjorken:1966jh,PhysRevD.1.1376}, which in QCD gets radiative corrections \cite{Baikov:2010je}, 
	\begin{eqnarray}
	\Gamma^{P-n}_\text{Bj}(Q^2) 
	=  \left|{g_A}/{g_V}\right|
	\left[1+C_{NS}(\alpha_s(Q^2))\right]/6.
	\label{E: Bj}
	\end{eqnarray}
	Using $C_{NS}(0.25) = -0.1195$ 
	and $\left|{g_A}/{g_V}\right|=1.2723\pm 0.0023$ 
	(measured in neutron beta decay) \cite{Patrignani:2016xqp} 
	gives $\Gamma_\text{Bj}^{P-n}(3~\mathrm{GeV}^2)=0.187$ (arrow in Fig.\ \ref{Fig: sumrules}) in very good agreement with the experimental value, $\Gamma^{P-n} =  0.181 \pm 0.008~ (\text{stat.}) \pm 0.014~ (\text{syst.})$  \cite{Adolph:2015saz}. 
	However, our model utilizing $\Delta f_{i/H}^\text{SU(6)}$ is in disagreement with \eqref{E: Bj}. This indicates a breaking of the naive $\mathrm{SU}(6)$ symmetry. 
	
	Therefore we abandon $\mathrm{SU}(6)$ symmetry for the nucleon and explore which values of $\Delta f_u^P$ and $\Delta f_d^P$ are required to reproduce the Bjorken sum rule value $\Gamma^{P- n}(0)$. Due to the probabilistic interpretation of the (bare) PDFs, these parameters cannot be varied freely because one has to ensure that the number of quarks with a specific spin orientation does not exceed the total number of quarks, i.e.\ $-2 \le  \Delta f_u^P \le 2$ and $-1 \le \Delta f_d^P \le 1$. 
	
	To reproduce the value of the integral $\Gamma^{P-n}(0)$ these parameters have to take their limiting values
	\begin{eqnarray} 
	\label{eq:final-delta-ud}
	\Delta f_u^P = 2 \,, \qquad \Delta f_d^P = -1   \,,
	\end{eqnarray}
	in striking contrast to the $\mathrm{SU}(6)$ values.

	With \eqref{eq:final-delta-ud} one obtains the full curves in Figs.\ \ref{figure:PDFS}-\ref{Fig: A1} 
	and the lower limit of the bands shown in Fig.\ \ref{Fig: sumrules}. Being still somewhat on the low side of the Bjorken sum rule, one should realize that a perfect agreement is not expected as long as our parameters are not fine-tuned to data. To demonstrate the potential of our model, we explore the impact of varying the parameter $\Lambda_H$ in (\ref{E: parameter-values}). Gradually reducing it by 10\% (which does not upset model agreement with unpolarized structure function data \cite{Ekstedt:2018onk}) leads to the bands in Fig.\ \ref{Fig: sumrules} giving a satisfying agreement with the Bjorken sum rule. This demonstrates that our model is also capable to describe the spin structure data. Interestingly, \eqref{eq:final-delta-ud} suggests in sharp contrast to the non-relativistic quark model that the overwhelming part of the up (down) quarks in the bare proton has spins parallel (opposite) to the proton spin. 

	Finally, the overall result in terms of the fraction of the proton spin carried by quarks is provided in Table \ref{Table: DeltaSigma}. The model result for $\Delta \Sigma$ is reduced and approaches the experimental value when more of the physically significant effects are included. The value does not change by adding the $\mathrm{SU}(6)$ breaking, due to $\Delta f_u^P+\Delta f_d^P$ being the same for the non-relativistic quark model and for \eqref{eq:final-delta-ud}. 
	The final value of $\Delta\Sigma$ compares reasonably well with the experimental result. 
	\begin{table}[t]\caption{$\Delta\Sigma$ at $Q^2 =3~\mathrm{GeV}^2$ from our model with bare proton, hadronic fluctuations, Melosh relativistic spin, SU(6) breaking compared to COMPASS data \cite{Adolph:2015saz}. } \label{Table: DeltaSigma}
		\begin{tabular}{c | c | c | c | c || c }
			\hline
			& bare  & hadronic fluct.\  & Melosh  & SU(6) break. & $\Delta \Sigma^\text{exp} $
			\\ \hline
			$\Delta \Sigma$  & 0.95 & 0.75 & $0.39$ & $0.39$ & $0.26$--$0.36$ 
		\end{tabular}
	\end{table} 
	%
	
	\section{Conclusions}\label{Sec: Conclusions}
	In a proton one observes, depending on the resolution, hadronic fluctuations, constituent partons and partonic fluctuations. Only the latter are described by perturbative QCD, while all ingredients are important for the description of the PDFs. A simple model that distributes the partons in a hadron with a statistical Gaussian weight, folded with the probabilities for specific hadronic fluctuations is capable of describing a large wealth of DIS structure function data. The input is fixed by low-energy hadron phenomenology except for the five parameters (\ref{E: parameter-values}) which take physically expected values when compared to data \cite{Ekstedt:2018onk}. With the choice \eqref{eq:final-delta-ud} that encodes the deviation from the spin-structure prediction of the non-relativistic quark model, the spin dependent structure functions can be added to the list of successfully described DIS data. It is encouraging that the well-established DGLAP equations of perturbative QCD, low-energy hadron physics and statistical distributions of many-body physics play so fruitfully together in the quest to understand the structure of the nucleon. 
	\newline\\~
	{\bf Acknowledgments}
	This work was supported by the Swedish Research Council under contract 621-2011-5107.
	
	\bibliographystyle{apsrev4-1}
	\bibliography{Spin.bib} 

\begin{thebibliography}{35}%
\makeatletter
\providecommand \@ifxundefined [1]{%
 \@ifx{#1\undefined}
}%
\providecommand \@ifnum [1]{%
 \ifnum #1\expandafter \@firstoftwo
 \else \expandafter \@secondoftwo
 \fi
}%
\providecommand \@ifx [1]{%
 \ifx #1\expandafter \@firstoftwo
 \else \expandafter \@secondoftwo
 \fi
}%
\providecommand \natexlab [1]{#1}%
\providecommand \enquote  [1]{``#1''}%
\providecommand \bibnamefont  [1]{#1}%
\providecommand \bibfnamefont [1]{#1}%
\providecommand \citenamefont [1]{#1}%
\providecommand \href@noop [0]{\@secondoftwo}%
\providecommand \href [0]{\begingroup \@sanitize@url \@href}%
\providecommand \@href[1]{\@@startlink{#1}\@@href}%
\providecommand \@@href[1]{\endgroup#1\@@endlink}%
\providecommand \@sanitize@url [0]{\catcode `\\12\catcode `\$12\catcode
  `\&12\catcode `\#12\catcode `\^12\catcode `\_12\catcode `\%12\relax}%
\providecommand \@@startlink[1]{}%
\providecommand \@@endlink[0]{}%
\providecommand \url  [0]{\begingroup\@sanitize@url \@url }%
\providecommand \@url [1]{\endgroup\@href {#1}{\urlprefix }}%
\providecommand \urlprefix  [0]{URL }%
\providecommand \Eprint [0]{\href }%
\providecommand \doibase [0]{http://dx.doi.org/}%
\providecommand \selectlanguage [0]{\@gobble}%
\providecommand \bibinfo  [0]{\@secondoftwo}%
\providecommand \bibfield  [0]{\@secondoftwo}%
\providecommand \translation [1]{[#1]}%
\providecommand \BibitemOpen [0]{}%
\providecommand \bibitemStop [0]{}%
\providecommand \bibitemNoStop [0]{.\EOS\space}%
\providecommand \EOS [0]{\spacefactor3000\relax}%
\providecommand \BibitemShut  [1]{\csname bibitem#1\endcsname}%
\let\auto@bib@innerbib\@empty
\bibitem [{\citenamefont {Ashman}\ \emph {et~al.}(1988)\citenamefont {Ashman}
  \emph {et~al.}}]{Ashman:1987hv}%
  \BibitemOpen
  \bibfield  {author} {\bibinfo {author} {\bibfnamefont {J.}~\bibnamefont
  {Ashman}} \emph {et~al.} (\bibinfo {collaboration} {European Muon
  Collaboration}),\ }\href {\doibase 10.1016/0370-2693(88)91523-7} {\bibfield
  {journal} {\bibinfo  {journal} {Phys. Lett.}\ }\textbf {\bibinfo {volume}
  {B206}},\ \bibinfo {pages} {364} (\bibinfo {year} {1988})}\BibitemShut
  {NoStop}%
\bibitem [{\citenamefont {Jaffe}\ and\ \citenamefont
  {Manohar}(1990)}]{JAFFE1990509}%
  \BibitemOpen
  \bibfield  {author} {\bibinfo {author} {\bibfnamefont {R.~L.}\ \bibnamefont
  {Jaffe}}\ and\ \bibinfo {author} {\bibfnamefont {A.}~\bibnamefont
  {Manohar}},\ }\href {\doibase 10.1016/0550-3213(90)90506-9} {\bibfield
  {journal} {\bibinfo  {journal} {Nucl. Phys.}\ }\textbf {\bibinfo {volume}
  {B337}},\ \bibinfo {pages} {509} (\bibinfo {year} {1990})}\BibitemShut
  {NoStop}%
\bibitem [{\citenamefont {Filippone}\ and\ \citenamefont
  {Ji}(2001)}]{Filippone:2001ux}%
  \BibitemOpen
  \bibfield  {author} {\bibinfo {author} {\bibfnamefont {B.~W.}\ \bibnamefont
  {Filippone}}\ and\ \bibinfo {author} {\bibfnamefont {X.-D.}\ \bibnamefont
  {Ji}},\ }\href {\doibase 10.1007/0-306-47915-X_1} {\bibfield  {journal}
  {\bibinfo  {journal} {Adv. Nucl. Phys.}\ }\textbf {\bibinfo {volume} {26}},\
  \bibinfo {pages} {1} (\bibinfo {year} {2001})},\ \Eprint
  {http://arxiv.org/abs/hep-ph/0101224} {arXiv:hep-ph/0101224} \BibitemShut
  {NoStop}%
\bibitem [{\citenamefont {Myhrer}\ and\ \citenamefont
  {Thomas}(2010)}]{Myhrer:2009uq}%
  \BibitemOpen
  \bibfield  {author} {\bibinfo {author} {\bibfnamefont {F.}~\bibnamefont
  {Myhrer}}\ and\ \bibinfo {author} {\bibfnamefont {A.~W.}\ \bibnamefont
  {Thomas}},\ }\href {\doibase 10.1088/0954-3899/37/2/023101} {\bibfield
  {journal} {\bibinfo  {journal} {J. Phys.}\ }\textbf {\bibinfo {volume}
  {G37}},\ \bibinfo {pages} {023101} (\bibinfo {year} {2010})},\ \Eprint
  {http://arxiv.org/abs/0911.1974} {arXiv:0911.1974} \BibitemShut {NoStop}%
\bibitem [{\citenamefont {Aidala}\ \emph {et~al.}(2013)\citenamefont {Aidala},
  \citenamefont {Bass}, \citenamefont {Hasch},\ and\ \citenamefont
  {Mallot}}]{Aidala:2012mv}%
  \BibitemOpen
  \bibfield  {author} {\bibinfo {author} {\bibfnamefont {C.~A.}\ \bibnamefont
  {Aidala}}, \bibinfo {author} {\bibfnamefont {S.~D.}\ \bibnamefont {Bass}},
  \bibinfo {author} {\bibfnamefont {D.}~\bibnamefont {Hasch}}, \ and\ \bibinfo
  {author} {\bibfnamefont {G.~K.}\ \bibnamefont {Mallot}},\ }\href {\doibase
  10.1103/RevModPhys.85.655} {\bibfield  {journal} {\bibinfo  {journal} {Rev.
  Mod. Phys.}\ }\textbf {\bibinfo {volume} {85}},\ \bibinfo {pages} {655}
  (\bibinfo {year} {2013})},\ \Eprint {http://arxiv.org/abs/1209.2803}
  {arXiv:1209.2803} \BibitemShut {NoStop}%
\bibitem [{\citenamefont {Bass}(2009)}]{Bass:2009dr}%
  \BibitemOpen
  \bibfield  {author} {\bibinfo {author} {\bibfnamefont {S.~D.}\ \bibnamefont
  {Bass}},\ }\href {\doibase 10.1142/S0217732309031041} {\bibfield  {journal}
  {\bibinfo  {journal} {Mod. Phys. Lett.}\ }\textbf {\bibinfo {volume} {A24}},\
  \bibinfo {pages} {1087} (\bibinfo {year} {2009})},\ \Eprint
  {http://arxiv.org/abs/0905.4619} {arXiv:0905.4619 [hep-ph]} \BibitemShut
  {NoStop}%
\bibitem [{\citenamefont {Adolph}\ \emph {et~al.}(2016)\citenamefont {Adolph}
  \emph {et~al.}}]{Adolph:2015saz}%
  \BibitemOpen
  \bibfield  {author} {\bibinfo {author} {\bibfnamefont {C.}~\bibnamefont
  {Adolph}} \emph {et~al.} (\bibinfo {collaboration} {COMPASS}),\ }\href
  {\doibase 10.1016/j.physletb.2015.11.064} {\bibfield  {journal} {\bibinfo
  {journal} {Phys. Lett.}\ }\textbf {\bibinfo {volume} {B753}},\ \bibinfo
  {pages} {18} (\bibinfo {year} {2016})},\ \Eprint
  {http://arxiv.org/abs/1503.08935} {arXiv:1503.08935} \BibitemShut {NoStop}%
\bibitem [{\citenamefont {Airapetian}\ \emph {et~al.}(2007)\citenamefont
  {Airapetian} \emph {et~al.}}]{Airapetian:2006vy}%
  \BibitemOpen
  \bibfield  {author} {\bibinfo {author} {\bibfnamefont {A.}~\bibnamefont
  {Airapetian}} \emph {et~al.} (\bibinfo {collaboration} {HERMES}),\ }\href
  {\doibase 10.1103/PhysRevD.75.012007} {\bibfield  {journal} {\bibinfo
  {journal} {Phys. Rev.}\ }\textbf {\bibinfo {volume} {D75}},\ \bibinfo {pages}
  {012007} (\bibinfo {year} {2007})},\ \Eprint
  {http://arxiv.org/abs/hep-ex/0609039} {arXiv:hep-ex/0609039} \BibitemShut
  {NoStop}%
\bibitem [{\citenamefont {Gribov}\ and\ \citenamefont
  {Lipatov}(1972)}]{Gribov:1972ri}%
  \BibitemOpen
  \bibfield  {author} {\bibinfo {author} {\bibfnamefont {V.~N.}\ \bibnamefont
  {Gribov}}\ and\ \bibinfo {author} {\bibfnamefont {L.~N.}\ \bibnamefont
  {Lipatov}},\ }\href@noop {} {\bibfield  {journal} {\bibinfo  {journal} {Sov.
  J. Nucl. Phys.}\ }\textbf {\bibinfo {volume} {15}},\ \bibinfo {pages} {438}
  (\bibinfo {year} {1972})},\ \bibinfo {note} {[Yad. Fiz. 15, 781
  (1972)]}\BibitemShut {NoStop}%
\bibitem [{\citenamefont {Dokshitzer}(1977)}]{Dokshitzer:1977sg}%
  \BibitemOpen
  \bibfield  {author} {\bibinfo {author} {\bibfnamefont {Y.~L.}\ \bibnamefont
  {Dokshitzer}},\ }\href@noop {} {\bibfield  {journal} {\bibinfo  {journal}
  {Sov. Phys. JETP}\ }\textbf {\bibinfo {volume} {46}},\ \bibinfo {pages} {641}
  (\bibinfo {year} {1977})},\ \bibinfo {note} {[Zh. Eksp. Teor. Fiz. 73, 1216
  (1977)]}\BibitemShut {NoStop}%
\bibitem [{\citenamefont {Altarelli}\ and\ \citenamefont
  {Parisi}(1977)}]{Altarelli:1977zs}%
  \BibitemOpen
  \bibfield  {author} {\bibinfo {author} {\bibfnamefont {G.}~\bibnamefont
  {Altarelli}}\ and\ \bibinfo {author} {\bibfnamefont {G.}~\bibnamefont
  {Parisi}},\ }\href {\doibase 10.1016/0550-3213(77)90384-4} {\bibfield
  {journal} {\bibinfo  {journal} {Nucl. Phys.}\ }\textbf {\bibinfo {volume}
  {B126}},\ \bibinfo {pages} {298} (\bibinfo {year} {1977})}\BibitemShut
  {NoStop}%
\bibitem [{\citenamefont {Ekstedt}\ \emph {et~al.}(2018)\citenamefont
  {Ekstedt}, \citenamefont {Ghaderi}, \citenamefont {Ingelman},\ and\
  \citenamefont {Leupold}}]{Ekstedt:2018onk}%
  \BibitemOpen
  \bibfield  {author} {\bibinfo {author} {\bibfnamefont {A.}~\bibnamefont
  {Ekstedt}}, \bibinfo {author} {\bibfnamefont {H.}~\bibnamefont {Ghaderi}},
  \bibinfo {author} {\bibfnamefont {G.}~\bibnamefont {Ingelman}}, \ and\
  \bibinfo {author} {\bibfnamefont {S.}~\bibnamefont {Leupold}},\ }\href@noop
  {} {\  (\bibinfo {year} {2018})},\ \Eprint {http://arxiv.org/abs/1807.06589}
  {arXiv:1807.06589} \BibitemShut {NoStop}%
\bibitem [{\citenamefont {Jenkins}\ and\ \citenamefont
  {Manohar}(1991)}]{Jenkins:1991es}%
  \BibitemOpen
  \bibfield  {author} {\bibinfo {author} {\bibfnamefont {E.~E.}\ \bibnamefont
  {Jenkins}}\ and\ \bibinfo {author} {\bibfnamefont {A.~V.}\ \bibnamefont
  {Manohar}},\ }\href {\doibase 10.1016/0370-2693(91)90840-M} {\bibfield
  {journal} {\bibinfo  {journal} {Phys. Lett.}\ }\textbf {\bibinfo {volume}
  {B259}},\ \bibinfo {pages} {353} (\bibinfo {year} {1991})}\BibitemShut
  {NoStop}%
\bibitem [{\citenamefont {Pascalutsa}\ and\ \citenamefont
  {Timmermans}(1999)}]{Pascalutsa:1999zz}%
  \BibitemOpen
  \bibfield  {author} {\bibinfo {author} {\bibfnamefont {V.}~\bibnamefont
  {Pascalutsa}}\ and\ \bibinfo {author} {\bibfnamefont {R.}~\bibnamefont
  {Timmermans}},\ }\href {\doibase 10.1103/PhysRevC.60.042201} {\bibfield
  {journal} {\bibinfo  {journal} {Phys. Rev.}\ }\textbf {\bibinfo {volume}
  {C60}},\ \bibinfo {pages} {042201(R)} (\bibinfo {year} {1999})},\ \Eprint
  {http://arxiv.org/abs/nucl-th/9905065} {arXiv:nucl-th/9905065} \BibitemShut
  {NoStop}%
\bibitem [{\citenamefont {Pascalutsa}\ \emph {et~al.}(2007)\citenamefont
  {Pascalutsa}, \citenamefont {Vanderhaeghen},\ and\ \citenamefont
  {Yang}}]{Pascalutsa:2006up}%
  \BibitemOpen
  \bibfield  {author} {\bibinfo {author} {\bibfnamefont {V.}~\bibnamefont
  {Pascalutsa}}, \bibinfo {author} {\bibfnamefont {M.}~\bibnamefont
  {Vanderhaeghen}}, \ and\ \bibinfo {author} {\bibfnamefont {S.~N.}\
  \bibnamefont {Yang}},\ }\href {\doibase 10.1016/j.physrep.2006.09.006}
  {\bibfield  {journal} {\bibinfo  {journal} {Phys. Rept.}\ }\textbf {\bibinfo
  {volume} {437}},\ \bibinfo {pages} {125} (\bibinfo {year} {2007})},\ \Eprint
  {http://arxiv.org/abs/hep-ph/0609004} {arXiv:hep-ph/0609004} \BibitemShut
  {NoStop}%
\bibitem [{\citenamefont {Ledwig}\ \emph {et~al.}(2014)\citenamefont {Ledwig},
  \citenamefont {Martin~Camalich}, \citenamefont {Geng},\ and\ \citenamefont
  {Vicente~Vacas}}]{Ledwig:2014rfa}%
  \BibitemOpen
  \bibfield  {author} {\bibinfo {author} {\bibfnamefont {T.}~\bibnamefont
  {Ledwig}}, \bibinfo {author} {\bibfnamefont {J.}~\bibnamefont
  {Martin~Camalich}}, \bibinfo {author} {\bibfnamefont {L.~S.}\ \bibnamefont
  {Geng}}, \ and\ \bibinfo {author} {\bibfnamefont {M.~J.}\ \bibnamefont
  {Vicente~Vacas}},\ }\href {\doibase 10.1103/PhysRevD.90.054502} {\bibfield
  {journal} {\bibinfo  {journal} {Phys. Rev.}\ }\textbf {\bibinfo {volume}
  {D90}},\ \bibinfo {pages} {054502} (\bibinfo {year} {2014})},\ \Eprint
  {http://arxiv.org/abs/1405.5456} {arXiv:1405.5456} \BibitemShut {NoStop}%
\bibitem [{\citenamefont {Arndt}\ and\ \citenamefont
  {Savage}(2002)}]{Arndt:2001ye}%
  \BibitemOpen
  \bibfield  {author} {\bibinfo {author} {\bibfnamefont {D.}~\bibnamefont
  {Arndt}}\ and\ \bibinfo {author} {\bibfnamefont {M.~J.}\ \bibnamefont
  {Savage}},\ }\href {\doibase 10.1016/S0375-9474(01)01223-4} {\bibfield
  {journal} {\bibinfo  {journal} {Nucl. Phys.}\ }\textbf {\bibinfo {volume}
  {A697}},\ \bibinfo {pages} {429} (\bibinfo {year} {2002})},\ \Eprint
  {http://arxiv.org/abs/nucl-th/0105045} {arXiv:nucl-th/0105045 [nucl-th]}
  \BibitemShut {NoStop}%
\bibitem [{\citenamefont {Chen}\ and\ \citenamefont {Ji}(2001)}]{Chen:2001eg}%
  \BibitemOpen
  \bibfield  {author} {\bibinfo {author} {\bibfnamefont {J.-W.}\ \bibnamefont
  {Chen}}\ and\ \bibinfo {author} {\bibfnamefont {X.-d.}\ \bibnamefont {Ji}},\
  }\href {\doibase 10.1016/S0370-2693(01)01337-5} {\bibfield  {journal}
  {\bibinfo  {journal} {Phys. Lett.}\ }\textbf {\bibinfo {volume} {B523}},\
  \bibinfo {pages} {107} (\bibinfo {year} {2001})},\ \Eprint
  {http://arxiv.org/abs/hep-ph/0105197} {arXiv:hep-ph/0105197 [hep-ph]}
  \BibitemShut {NoStop}%
\bibitem [{\citenamefont {Chen}\ and\ \citenamefont {Ji}(2002)}]{Chen:2001pva}%
  \BibitemOpen
  \bibfield  {author} {\bibinfo {author} {\bibfnamefont {J.-W.}\ \bibnamefont
  {Chen}}\ and\ \bibinfo {author} {\bibfnamefont {X.-d.}\ \bibnamefont {Ji}},\
  }\href {\doibase 10.1103/PhysRevLett.88.052003} {\bibfield  {journal}
  {\bibinfo  {journal} {Phys. Rev. Lett.}\ }\textbf {\bibinfo {volume} {88}},\
  \bibinfo {pages} {052003} (\bibinfo {year} {2002})},\ \Eprint
  {http://arxiv.org/abs/hep-ph/0111048} {arXiv:hep-ph/0111048 [hep-ph]}
  \BibitemShut {NoStop}%
\bibitem [{\citenamefont {Thomas}\ \emph {et~al.}(2000)\citenamefont {Thomas},
  \citenamefont {Melnitchouk},\ and\ \citenamefont {Steffens}}]{Thomas:2000ny}%
  \BibitemOpen
  \bibfield  {author} {\bibinfo {author} {\bibfnamefont {A.~W.}\ \bibnamefont
  {Thomas}}, \bibinfo {author} {\bibfnamefont {W.}~\bibnamefont {Melnitchouk}},
  \ and\ \bibinfo {author} {\bibfnamefont {F.~M.}\ \bibnamefont {Steffens}},\
  }\href {\doibase 10.1103/PhysRevLett.85.2892} {\bibfield  {journal} {\bibinfo
   {journal} {Phys. Rev. Lett.}\ }\textbf {\bibinfo {volume} {85}},\ \bibinfo
  {pages} {2892} (\bibinfo {year} {2000})},\ \Eprint
  {http://arxiv.org/abs/hep-ph/0005043} {arXiv:hep-ph/0005043 [hep-ph]}
  \BibitemShut {NoStop}%
\bibitem [{\citenamefont {Ji}\ \emph {et~al.}(2013)\citenamefont {Ji},
  \citenamefont {Melnitchouk},\ and\ \citenamefont {Thomas}}]{Ji:2013bca}%
  \BibitemOpen
  \bibfield  {author} {\bibinfo {author} {\bibfnamefont {C.-R.}\ \bibnamefont
  {Ji}}, \bibinfo {author} {\bibfnamefont {W.}~\bibnamefont {Melnitchouk}}, \
  and\ \bibinfo {author} {\bibfnamefont {A.~W.}\ \bibnamefont {Thomas}},\
  }\href {\doibase 10.1103/PhysRevD.88.076005} {\bibfield  {journal} {\bibinfo
  {journal} {Phys. Rev.}\ }\textbf {\bibinfo {volume} {D88}},\ \bibinfo {pages}
  {076005} (\bibinfo {year} {2013})},\ \Eprint {http://arxiv.org/abs/1306.6073}
  {arXiv:1306.6073 [hep-ph]} \BibitemShut {NoStop}%
\bibitem [{\citenamefont {Wang}\ \emph {et~al.}(2016)\citenamefont {Wang},
  \citenamefont {Ji}, \citenamefont {Melnitchouk}, \citenamefont {Salamu},
  \citenamefont {Thomas},\ and\ \citenamefont {Wang}}]{Wang:2016ndh}%
  \BibitemOpen
  \bibfield  {author} {\bibinfo {author} {\bibfnamefont {X.~G.}\ \bibnamefont
  {Wang}}, \bibinfo {author} {\bibfnamefont {C.-R.}\ \bibnamefont {Ji}},
  \bibinfo {author} {\bibfnamefont {W.}~\bibnamefont {Melnitchouk}}, \bibinfo
  {author} {\bibfnamefont {Y.}~\bibnamefont {Salamu}}, \bibinfo {author}
  {\bibfnamefont {A.~W.}\ \bibnamefont {Thomas}}, \ and\ \bibinfo {author}
  {\bibfnamefont {P.}~\bibnamefont {Wang}},\ }\href {\doibase
  10.1103/PhysRevD.94.094035} {\bibfield  {journal} {\bibinfo  {journal} {Phys.
  Rev.}\ }\textbf {\bibinfo {volume} {D94}},\ \bibinfo {pages} {094035}
  (\bibinfo {year} {2016})},\ \Eprint {http://arxiv.org/abs/1610.03333}
  {arXiv:1610.03333 [hep-ph]} \BibitemShut {NoStop}%
\bibitem [{\citenamefont {Botje}(2011)}]{Botje:2010ay}%
  \BibitemOpen
  \bibfield  {author} {\bibinfo {author} {\bibfnamefont {M.}~\bibnamefont
  {Botje}},\ }\href {\doibase 10.1016/j.cpc.2010.10.020} {\bibfield  {journal}
  {\bibinfo  {journal} {Comput. Phys. Commun.}\ }\textbf {\bibinfo {volume}
  {182}},\ \bibinfo {pages} {490} (\bibinfo {year} {2011})},\ \Eprint
  {http://arxiv.org/abs/1005.1481} {arXiv:1005.1481} \BibitemShut {NoStop}%
\bibitem [{\citenamefont {Mosel}(1999)}]{mosel1999fields}%
  \BibitemOpen
  \bibfield  {author} {\bibinfo {author} {\bibfnamefont {U.}~\bibnamefont
  {Mosel}},\ }\href {https://books.google.se/books?id=bWQNYnMj\_7AC} {\emph
  {\bibinfo {title} {Fields, Symmetries, and Quarks}}},\ Theoretical and
  Mathematical Physics\ (\bibinfo  {publisher} {Springer Berlin Heidelberg},\
  \bibinfo {year} {1999})\BibitemShut {NoStop}%
\bibitem [{\citenamefont {Melosh}(1974)}]{Melosh:1974cu}%
  \BibitemOpen
  \bibfield  {author} {\bibinfo {author} {\bibfnamefont {H.~J.}\ \bibnamefont
  {Melosh}},\ }\href {\doibase 10.1103/PhysRevD.9.1095} {\bibfield  {journal}
  {\bibinfo  {journal} {Phys. Rev.}\ }\textbf {\bibinfo {volume} {D9}},\
  \bibinfo {pages} {1095} (\bibinfo {year} {1974})}\BibitemShut {NoStop}%
\bibitem [{\citenamefont {Ma}(1991)}]{Ma:1991xq}%
  \BibitemOpen
  \bibfield  {author} {\bibinfo {author} {\bibfnamefont {B.-Q.}\ \bibnamefont
  {Ma}},\ }\href {\doibase 10.1088/0954-3899/17/5/001} {\bibfield  {journal}
  {\bibinfo  {journal} {J. Phys.}\ }\textbf {\bibinfo {volume} {G17}},\
  \bibinfo {pages} {L53} (\bibinfo {year} {1991})},\ \Eprint
  {http://arxiv.org/abs/0711.2335} {arXiv:0711.2335} \BibitemShut {NoStop}%
\bibitem [{\citenamefont {Beyer}\ \emph {et~al.}(1998)\citenamefont {Beyer},
  \citenamefont {Kuhrts},\ and\ \citenamefont {Weber}}]{Beyer:1998xy}%
  \BibitemOpen
  \bibfield  {author} {\bibinfo {author} {\bibfnamefont {M.}~\bibnamefont
  {Beyer}}, \bibinfo {author} {\bibfnamefont {C.}~\bibnamefont {Kuhrts}}, \
  and\ \bibinfo {author} {\bibfnamefont {H.~J.}\ \bibnamefont {Weber}},\ }\href
  {\doibase 10.1006/aphy.1998.5837} {\bibfield  {journal} {\bibinfo  {journal}
  {Annals Phys.}\ }\textbf {\bibinfo {volume} {269}},\ \bibinfo {pages} {129}
  (\bibinfo {year} {1998})},\ \Eprint {http://arxiv.org/abs/nucl-th/9804021}
  {arXiv:nucl-th/9804021} \BibitemShut {NoStop}%
\bibitem [{\citenamefont {Sun}\ and\ \citenamefont {Weber}(2002)}]{Sun:2001ir}%
  \BibitemOpen
  \bibfield  {author} {\bibinfo {author} {\bibfnamefont {X.-p.}\ \bibnamefont
  {Sun}}\ and\ \bibinfo {author} {\bibfnamefont {H.~J.}\ \bibnamefont
  {Weber}},\ }\href {\doibase 10.1142/S0217751X02009710} {\bibfield  {journal}
  {\bibinfo  {journal} {Int. J. Mod. Phys.}\ }\textbf {\bibinfo {volume}
  {A17}},\ \bibinfo {pages} {2535} (\bibinfo {year} {2002})},\ \Eprint
  {http://arxiv.org/abs/hep-ph/0102240} {arXiv:hep-ph/0102240} \BibitemShut
  {NoStop}%
\bibitem [{\citenamefont {Schmidt}\ and\ \citenamefont
  {Soffer}(1997)}]{SCHMIDT1997331}%
  \BibitemOpen
  \bibfield  {author} {\bibinfo {author} {\bibfnamefont {I.}~\bibnamefont
  {Schmidt}}\ and\ \bibinfo {author} {\bibfnamefont {J.}~\bibnamefont
  {Soffer}},\ }\href {\doibase 10.1016/S0370-2693(97)00737-5} {\bibfield
  {journal} {\bibinfo  {journal} {Phys. Lett.}\ }\textbf {\bibinfo {volume}
  {B407}},\ \bibinfo {pages} {331} (\bibinfo {year} {1997})},\ \Eprint
  {http://arxiv.org/abs/hep-ph/9703411} {arXiv:hep-ph/9703411} \BibitemShut
  {NoStop}%
\bibitem [{\citenamefont {Ellis}\ and\ \citenamefont
  {Jaffe}(1974)}]{Ellis:1973kp}%
  \BibitemOpen
  \bibfield  {author} {\bibinfo {author} {\bibfnamefont {J.~R.}\ \bibnamefont
  {Ellis}}\ and\ \bibinfo {author} {\bibfnamefont {R.~L.}\ \bibnamefont
  {Jaffe}},\ }\href {\doibase 10.1103/physrevd.10.1669.4,
  10.1103/PhysRevD.9.1444} {\bibfield  {journal} {\bibinfo  {journal} {Phys.
  Rev.}\ }\textbf {\bibinfo {volume} {D9}},\ \bibinfo {pages} {1444} (\bibinfo
  {year} {1974})},\ \bibinfo {note} {[Erratum: Phys. Rev. {\bf D10}, 1669
  (1974)]}\BibitemShut {NoStop}%
\bibitem [{\citenamefont {Bjorken}(1966)}]{Bjorken:1966jh}%
  \BibitemOpen
  \bibfield  {author} {\bibinfo {author} {\bibfnamefont {J.~D.}\ \bibnamefont
  {Bjorken}},\ }\href {\doibase 10.1103/PhysRev.148.1467} {\bibfield  {journal}
  {\bibinfo  {journal} {Phys. Rev.}\ }\textbf {\bibinfo {volume} {148}},\
  \bibinfo {pages} {1467} (\bibinfo {year} {1966})}\BibitemShut {NoStop}%
\bibitem [{\citenamefont {Bjorken}(1970)}]{PhysRevD.1.1376}%
  \BibitemOpen
  \bibfield  {author} {\bibinfo {author} {\bibfnamefont {J.~D.}\ \bibnamefont
  {Bjorken}},\ }\href {\doibase 10.1103/PhysRevD.1.1376} {\bibfield  {journal}
  {\bibinfo  {journal} {Phys. Rev.}\ }\textbf {\bibinfo {volume} {D1}},\
  \bibinfo {pages} {1376} (\bibinfo {year} {1970})}\BibitemShut {NoStop}%
\bibitem [{\citenamefont {Alexakhin}\ \emph {et~al.}(2007)\citenamefont
  {Alexakhin} \emph {et~al.}}]{Alexakhin:2006oza}%
  \BibitemOpen
  \bibfield  {author} {\bibinfo {author} {\bibfnamefont {V.~{\relax Yu}.}\
  \bibnamefont {Alexakhin}} \emph {et~al.} (\bibinfo {collaboration}
  {COMPASS}),\ }\href {\doibase 10.1016/j.physletb.2006.12.076} {\bibfield
  {journal} {\bibinfo  {journal} {Phys. Lett.}\ }\textbf {\bibinfo {volume}
  {B647}},\ \bibinfo {pages} {8} (\bibinfo {year} {2007})},\ \Eprint
  {http://arxiv.org/abs/hep-ex/0609038} {arXiv:hep-ex/0609038} \BibitemShut
  {NoStop}%
\bibitem [{\citenamefont {Baikov}\ \emph {et~al.}(2010)\citenamefont {Baikov},
  \citenamefont {Chetyrkin},\ and\ \citenamefont {{K\"uhn}}}]{Baikov:2010je}%
  \BibitemOpen
  \bibfield  {author} {\bibinfo {author} {\bibfnamefont {P.~A.}\ \bibnamefont
  {Baikov}}, \bibinfo {author} {\bibfnamefont {K.~G.}\ \bibnamefont
  {Chetyrkin}}, \ and\ \bibinfo {author} {\bibfnamefont {J.~H.}\ \bibnamefont
  {{K\"uhn}}},\ }\href {\doibase 10.1103/PhysRevLett.104.132004} {\bibfield
  {journal} {\bibinfo  {journal} {Phys. Rev. Lett.}\ }\textbf {\bibinfo
  {volume} {104}},\ \bibinfo {pages} {132004} (\bibinfo {year} {2010})},\
  \Eprint {http://arxiv.org/abs/1001.3606} {arXiv:1001.3606} \BibitemShut
  {NoStop}%
\bibitem [{\citenamefont {Patrignani}\ \emph {et~al.}(2016)\citenamefont
  {Patrignani} \emph {et~al.}}]{Patrignani:2016xqp}%
  \BibitemOpen
  \bibfield  {author} {\bibinfo {author} {\bibfnamefont {C.}~\bibnamefont
  {Patrignani}} \emph {et~al.} (\bibinfo {collaboration} {Particle Data
  Group}),\ }\href {\doibase 10.1088/1674-1137/40/10/100001} {\bibfield
  {journal} {\bibinfo  {journal} {Chin. Phys.}\ }\textbf {\bibinfo {volume}
  {C40}},\ \bibinfo {pages} {100001} (\bibinfo {year} {2016})}\BibitemShut
  {NoStop}%
\end{thebibliography}%
\end{document}